# Spin Seebeck effect in quantum magnet Pb$_2$V$_3$O$_9$

Wenyu Xing[1]*, Ranran Cai[1], Kodai Moriyama[2], Kensuke Nara[2], Yunyan Yao[1], Weiliang Qiao[1], Kazuyoshi Yoshimura[2]*, and Wei Han[1]*

[1] International Center for Quantum Materials, School of Physics, Peking University, Beijing, 100871, P. R. China

[2] Department of Chemistry, Graduate School of Science, Kyoto University, Kyoto, 606-8502, Japan

*Correspondence to: wenyuxing@pku.edu.cn (W.X.);

yoshimura.kazuyoshi.8e@kyoto-u.ac.jp (K.Y.);

weihan@pku.edu.cn (W.H.)

**Abstract**

Spin Seebeck effect (SSE), the generation of spin current from heat, has been extensively studied in a large variety of magnetic materials, including ferromagnets, antiferromagnets, paramagnets, and quantum spin liquids. In this paper, we report the study of the SSE in the single crystalline Pb$_2$V$_3$O$_9$, a spin-gapped quantum magnet candidate with quasi-one-dimensional spin-1/2 chain. Detailed temperature and magnetic field dependences of the SSE are investigated, and the temperature-dependent critical magnetic fields show a strong correlation to the Bose-Einstein condensation phase of the quantum magnet Pb$_2$V$_3$O$_9$. This work shows the potential of using spin current as a probe to study the spin correlation and phase transition properties in quantum magnets.

Spin Seebeck effect (SSE) refers to the spin current generation by a temperature gradient, which has generated a lot of interests and leads to the rise of spin-caloritronics [1,2]. Physically, SSE is a spin-thermoelectric effect, which arises from the direct coupling between spin and heat [2-4]. Technically, it might be useful for clean energy harvesting from waste heat [5,6]. Due to its physical importance and potential technical impact, SSE has been extensively studied in a large variety of magnetic materials, including ferromagnets, ferrimagnets, and antiferromagnets [7-14]. Beyond the conventional magnetic materials, SSE has also been studied in the exotic magnets with strong quantum and thermal fluctuations, including geometrically frustrated magnets [15,16], quantum spin liquids [17,18], and spin-nematic Tomonaga-Luttinger liquids [19]. Furthermore, the SSE-generated spin current can be used to investigate the magnetic phase transition in antiferromagnetic insulators [20], and to study unusual spin-dependent physical properties in various quantum materials [21].

Quantum magnets are one kind of emergent quantum materials that exhibit Bose-Einstein condensation (BEC) [22-24]. When an external magnetic field ($B$) is applied, a triplon BEC is formed as the spin-singlet ground state gap is closed. $Pb_2V_3O_9$ (PVO) is a quantum magnet candidate with spin-1/2 dimers, which has a relatively low critical magnetic field to form the BEC states [25-30]. This BEC-related phase diagram has been investigated by numerous techniques, such as specific heat and torque measurements [28,29]. Since the BEC is the most basic ground state of a quantum magnet, the research on it might lead to the identification of the novel bosonic phases, such as Bose metal and Bose glass, in the quantum magnets [23].

In this letter, we report the observation of the SSE in the single crystalline quantum magnet PVO using the longitudinal geometry. Detailed temperature and magnetic field dependences of the SSE are investigated. Interestingly, at higher temperature, the SSE-generated spin current is mainly mediated by paramagnons, and can be enhanced by the external magnetic field. However, at lower temperature, SSE first increases, and then decreases as $B$ increases. The suppression of the SSE signal exhibits a strong correlation to the magnetic field-induced BEC phase, and the obtained magnetic phase diagram agrees well with previous reports. Our work demonstrates that SSE could be a unique probe of the spin correlation and phase transition in quantum magnets.

The PVO single crystals were synthesized using the standard floating zone method, and the details were described in a previous report [28]. Firstly, the stoichiometric PbO, $V_2O_5$, and $V_2O_3$ powders were mixed, and polycrystalline samples of PVO were formed via solid state reaction. Then, the polycrystalline PVO samples were transformed to single crystals using the floating zone method. The crystalline structure and the phase purity of the synthesized PVO single crystals were characterized by X-ray diffraction (D8 Advance, Bruker). The magnetization of single crystalline PVO was characterized by the superconducting quantum interference device in a magnetic properties measurement system (MPMS; Quantum Design).

The SSE in PVO was investigated in longitudinal SSE geometry, and the devices were fabricated as follows. Firstly, the surface of PVO single crystal was polished by lapping films (261X, 3M). Then the longitudinal SSE devices were fabricated on the flat surface using standard E-beam lithography and lift-off processes. The first step was to define a Pt electrode (width: 10 μm, length: 600 μm) on the surface of PVO via E-beam lithography. The 10-nm thick Pt was deposited in a magneton sputtering system with a base pressure lower than $8.0 \times 10^{-7}$ mbar. Then a second E-beam lithography step was used to define the heater on the top of the Pt electrode, which consists of a 100-nm-thick insulating $Al_2O_3$ layer and a 50-nm-thick Ti layer grown via E-beam evaporation.

The SSE measurement was performed via on-chip local heating method, and the SSE voltages were measured using the standard low-frequency lock-in technique in a physical properties measurement system (Dynacool, Quantum Design). The higher magnetic ordering temperature of 4 K under applied magnetic fields and the larger critical magnetic field of 3.5 T for $Pb_2V_3O_9$ make it possible to study SSE in the measurement system (Largest $B$ = 9 T) [23]. The AC current ($f$ = 7 Hz) was applied in the Ti electrode via a current source (K6221, Keithley) to generate the temperature gradient ($\nabla T$) perpendicular to the PVO surface. The SSE voltages were obtained via the second-harmonic voltage ($V_{2\omega}$) using lock-in amplifiers (SR830, Stanford Research) with a preamplifier (SR560, Stanford Research) for a better signal-to-noise ratio.

Figure 1(a) illustrates the crystal structure of PVO that belongs to the space group $C\bar{1}$. The magnetic $V^{4+}$ ions (spin angular momentum; $S$ = 1/2) construct the spin chains while the nonmagnetic $V^{5+}O_4$ tetrahedrons share the corners with the $V^{4+}O_6$ octahedrons [25,28,31]. An optical image of the typical PVO single crystal is shown in Fig. 1(b) inset. The single crystalline feature of the PVO crystal is confirmed via the X-ray diffraction. As shown in Fig. 1(b), only the (020)-oriented peaks are observed, which is consistent with the previous report [28]. Figure 1(c) shows the temperature dependence of magnetization under the in-plane magnetic field of 1 T. The magnetization shows a maximum around 20 K and then decreases rapidly as the temperature decreases down to 2 K. When the magnetic field increases, i.e. from 5.6 T to 7.0 T, an upturn of magnetization is observed at the low temperatures (Fig. 1(d)). The critical temperature $T_c$, indicated by arrows in Fig. 1(d), represents the magnetic field induced BEC phase transitions as demonstrated in previous reports [25,32-34].

Figure 2(a) shows the schematic of the fabricated SSE device in the longitudinal geometry, where the temperature gradient and the spin current are both perpendicular to the sample surface. For the SSE device, the Ti layer is used to generate the heat flow arising from Joule heating, and the $Al_2O_3$ layer is used for electric insulation between the Ti and Pt electrodes. Under the temperature gradient, a spin current ($J_s$) is generated at the PVO/Pt interface, and measured via inverse spin Hall effect in Pt by the voltage probes at the two ends of the Pt electrode. During the SSE measurement, the PVO device is rotated from 0° to 360° in a constant in-plane magnetic field, as illustrated in Fig. 2(b) inset. Figure 2(b) shows a typical $V_{2\omega}$ curve as a function of the magnetic field angle ($\varphi$) measured at $T = 2$ K and $B = 5$ T. $V_{2\omega}$ reaches maximum (minimum) when the magnetic field angle is 90° (270°) and the generated spin-current polarization is perpendicular to the Pt electrode. The measured $V_{2\omega}$ is proportional to sin ($\varphi$) [35]:

$$V_{2\omega} = V_{SSE} \sin(\varphi) \tag{1}$$

where $V_{SSE}$ is the SSE signal. The red solid line in Fig. 2(b) is the best-fitting curve based on Eq. (1), from which $V_{SSE}$ is determined to be 54±0.7 nV. The measured $V_{SSE}$ scales quadratically with the current ($I$) in the heater electrode, as shown in the right inset of Fig. 2(b), which is consistent with the expectations for SSE measurements.

Next, the temperature dependence of SSE in the single crystal of PVO is investigated. Figures 3(a) and 3(b) show the typical magnetic field angle dependences of $V_{2\omega}$ at $T = 7.5$ K and 20 K, which can both be well fitted by Eq. (1). The $V_{SSE}$ are determined to be 82±0.9 and 32±0.7 nV for $T = 7.5$ K and 20 K, respectively. Figure 3(c) summarizes the $V_{SSE}$ as a function of the temperature measured under $B = 5$ T. As $T$ decreases from 60 K to 2 K, $V_{SSE}$ first increases, reaches a maximum around $T = 7.5$ K, and then decreases rapidly. The temperature-dependent $V_{SSE}$ results could be attributed to the competition between the decreased paramagnon density and the increased lifetime as the temperature decreases [36]. Between 60 and 7.5 K, the lifetimes increase faster compared to the reduction of the paramagnon density, giving rise to an enhanced $V_{SSE}$ signal. But below $T = 7.5$ K, the reduction of the paramagnon density is more profound than the increase of the lifetimes, leading to a reduced $V_{SSE}$ signal. The temperature-dependent $V_{SSE}$ curves under $B = 1$, 4 and 9 T show very similar features, as shown in the inset of Fig. 3(c).

Figure 4(a) shows the $V_{SSE}$ as a function of magnetic field at $T = 2$, 2.5, 3, 4, 5, and 7.5 K, respectively. Under lower magnetic fields, $V_{SSE}$ increases as $B$ increases for all the temperatures. This behavior agrees well with expectations, since in paramagnets, magnetic field can enhance the SSE via elevating the dynamical spin susceptibility [13,19]. Under higher magnetic fields, $V_{SSE}$ exhibits different features. For higher temperatures ($T = 4, 5, 7.5$ K), $V_{SSE}$ is enhanced as the magnetic field increases up to 9 T. However, for lower temperatures, $V_{SSE}$ shows a decrease as the magnetic field increases under higher magnetic fields. The reduction of $V_{SSE}$ becomes more prominent with decreasing temperature. Figures 4(b-c) show the normalized SSE signal ($V_{SSE}/V_{max}$) curves. Obviously, the critical magnetic field ($B_c$; at which the maximum $V_{SSE}$ is observed) increases as the temperature increases, as indicated by the arrows. $B_c$ are determined to be $5.0 \pm 0.5$ T and $6.1 \pm 0.4$ T at $T = 2$ and 2.5 K, respectively. We speculate the reason for the reduction of $V_{SSE}$ is correlated to the magnon pair formation, as reported in a quasi-one-dimensional frustrated spin-1/2 magnet LiCuVO$_4$ [19]. The spin angular momentum transfer at the interface for one magnon requires two electrons in the Pt, but this process for a magnon pair requires four electrons in Pt. Hence, the efficiency of spin angular momentum transfer at the interface for quantum magnet BEC states might be less than the paramagnetic states. Nevertheless, this speculation requires further theoretical studies.

Based on the magnetic field dependence of SSE, all the values of $B_c$ are obtained from 2 K to 3.5 K, and the results are plotted in Fig. 4(d). Interestingly, these critical magnetic fields are similar to the values that are correlated to the BEC phase transition of PVO obtained in previous reports [25,26,28,29]. To quantitatively investigate this feature, the results of $T$ vs. $B_C$ are fitted by the following power law relationship that describes the BEC phase boundary around the lower critical field regime [22]:

$$T \propto (B - B_c)^{2/3} \quad (2)$$

Clearly, the experimental results and the power law-fitting curve match well with each other. Furthermore, the values of $B_c$ obtained from SSE measurement coincide with phase boundary determined by the present magnetization and previous specific heat and torque measurements [28,29], as shown in Fig. 4(d) inset. This demonstrates that the spin current generated by SSE can be a unique method to investigate the interesting spin-dependent transport properties and the related spin correlations in quantum magnets.

In summary, we report the temperature- and magnetic field-dependent SSE in the quantum magnet PVO using on-chip local heating method. At lower temperatures, a non-monotonic behavior of the magnetic field-dependent $V_{SSE}$ is observed. The temperature dependence of the critical magnetic field exhibits a strong correlation to the BEC phase of the quantum magnet PVO. Our work demonstrates that SSE could be a unique probe of the spin correlation and phase transition in quantum magnets.

**Acknowledgement**

This work is supported by National Basic Research Programs of China (2019YFA0308401), National Natural Science Foundation of China (11974025), and the Strategic Priority Research Program of the Chinese Academy of Sciences (No. XDB28000000).

**Data availability**

The data that support the findings of this study are available from the corresponding author upon reasonable request.

**References:**

[1] K. Uchida, M. Ishida, T. Kikkawa, A. Kirihara, T. Murakami, and E. Saitoh, *J. Phys.: Condens. Matter.* **26**, 343202 (2014).
[2] G. E. W. Bauer, E. Saitoh, and B. J. van Wees, *Nat. Mater.* **11**, 391 (2012).
[3] K. Uchida, S. Takahashi, K. Harii, J. Ieda, W. Koshibae, K. Ando, S. Maekawa, and E. Saitoh, *Nature* **455**, 778 (2008).
[4] K. Uchida, J. Xiao, H. Adachi *et al.*, *Nat. Mater.* **9**, 894 (2010).
[5] A. Kirihara, K.-i. Uchida, Y. Kajiwara, M. Ishida, Y. Nakamura, T. Manako, E. Saitoh, and S. Yorozu, *Nat. Mater.* **11**, 686 (2012).
[6] R. Ramos, T. Kikkawa, M. H. Aguirre *et al.*, *Phys. Rev. B* **92**, 220407 (2015).
[7] K. Uchida, H. Adachi, T. Ota, H. Nakayama, S. Maekawa, and E. Saitoh, *Appl. Phys. Lett.* **97**, 172505 (2010).
[8] K. Uchida, T. Nonaka, T. Kikkawa, Y. Kajiwara, and E. Saitoh, *Phys. Rev. B* **87**, 104412 (2013).
[9] S. Seki, T. Ideue, M. Kubota, Y. Kozuka, R. Takagi, M. Nakamura, Y. Kaneko, M. Kawasaki, and Y. Tokura, *Phys. Rev. Lett.* **115**, 266601 (2015).
[10] S. Geprägs, A. Kehlberger, F. D. Coletta *et al.*, *Nat. Commun.* **7**, 10452 (2016).
[11] S. M. Wu, W. Zhang, A. Kc, P. Borisov, J. E. Pearson, J. S. Jiang, D. Lederman, A. Hoffmann, and A. Bhattacharya, *Phys. Rev. Lett.* **116**, 097204 (2016).
[12] J. Shan, A. V. Singh, L. Liang, L. J. Cornelissen, Z. Galazka, A. Gupta, B. J. van Wees, and T. Kuschel, *Appl. Phys. Lett.* **113**, 162403 (2018).
[13] J. Li, Z. Shi, V. H. Ortiz, M. Aldosary, C. Chen, V. Aji, P. Wei, and J. Shi, *Phys. Rev. Lett.* **122**, 217204 (2019).
[14] W. Xing, Y. Ma, Y. Yao, R. Cai, Y. Ji, R. Xiong, K. Shen, and W. Han, *Phys. Rev. B* **102**, 184416 (2020).
[15] S. M. Wu, J. E. Pearson, and A. Bhattacharya, *Phys. Rev. Lett.* **114**, 186602 (2015).
[16] C. Liu, S. M. Wu, J. E. Pearson, J. S. Jiang, N. d'Ambrumenil, and A. Bhattacharya, *Phys. Rev. B* **98**, 060415 (2018).
[17] D. Hirobe, M. Sato, T. Kawamata, Y. Shiomi, K. Uchida, R. Iguchi, Y. Koike, S. Maekawa, and E. Saitoh, *Nat. Phys.* **13**, 30 (2016).
[18] D. Hirobe, T. Kawamata, K. Oyanagi, Y. Koike, and E. Saitoh, *J. Appl. Phys.* **123**, 123903 (2018).
[19] D. Hirobe, M. Sato, M. Hagihala, Y. Shiomi, T. Masuda, and E. Saitoh, *Phys. Rev. Lett.* **123**, 117202 (2019).
[20] Z. Qiu, D. Hou, J. Barker, K. Yamamoto, O. Gomonay, and E. Saitoh, *Nat. Mater.* **17**, 577 (2018).
[21] W. Han, S. Maekawa, and X.-C. Xie, *Nat. Mater.* **19**, 139 (2020).
[22] T. Giamarchi, C. Rüegg, and O. Tchernyshyov, *Nat. Phys.* **4**, 198 (2008).
[23] V. Zapf, M. Jaime, and C. D. Batista, *Rev. Mod. Phys.* **86**, 563 (2014).
[24] A. Vasiliev, O. Volkova, E. Zvereva, and M. Markina, *npj Quant. Mater.* **3**, 18 (2018).

[25] T. Waki, Y. Morimoto, C. Michioka *et al.*, *J. Phys. Soc. Jpn.* **73**, 3435 (2004).

[26] T. Waki, M. Kato, Y. Itoh, C. Michioka, K. Yoshimura, and T. Goto, *J. Phys. Chem. Solids* **66**, 1432 (2005).

[27] T. Waki, Y. Itoh, C. Michioka, K. Yoshimura, and M. Kato, *Phys. Rev. B* **73**, 064419 (2006).

[28] K. Nawa, C. Michioka, K. Yoshimura, A. Matsuo, and K. Kindo, *J. Phys. Soc. Jpn.* **80**, 034710 (2011).

[29] B. S. Conner, H. D. Zhou, Y. J. Jo, L. Balicas, C. R. Wiebe, J. P. Carlo, Y. J. Uemura, A. A. Aczel, T. J. Williams, and G. M. Luke, *Phys. Rev. B* **81**, 132401 (2010).

[30] A. A. Tsirlin and H. Rosner, *Phys. Rev. B* **83**, 064415 (2011).

[31] O. Mentré, A. C. Dhaussy, F. Abraham, E. Suard, and H. Steinfink, *Chem. Mater.* **11**, 2408 (1999).

[32] T. Nikuni, M. Oshikawa, A. Oosawa, and H. Tanaka, *Phys. Rev. Lett.* **84**, 5868 (2000).

[33] K. Shirasawa, N. Kurita, and H. Tanaka, *Phys. Rev. B* **96**, 144404 (2017).

[34] A. Paduan-Filho, X. Gratens, and N. F. Oliveira, *Phys. Rev. B* **69**, 020405 (2004).

[35] A. Aqeel, N. Vlietstra, J. A. Heuver, G. E. W. Bauer, B. Noheda, B. J. van Wees, and T. T. M. Palstra, *Phys. Rev. B* **92**, 224410 (2015).

[36] S. M. Rezende, R. L. Rodríguez-Suárez, R. O. Cunha, A. R. Rodrigues, F. L. A. Machado, G. A. Fonseca Guerra, J. C. Lopez Ortiz, and A. Azevedo, *Phys. Rev. B* **89**, 014416 (2014).

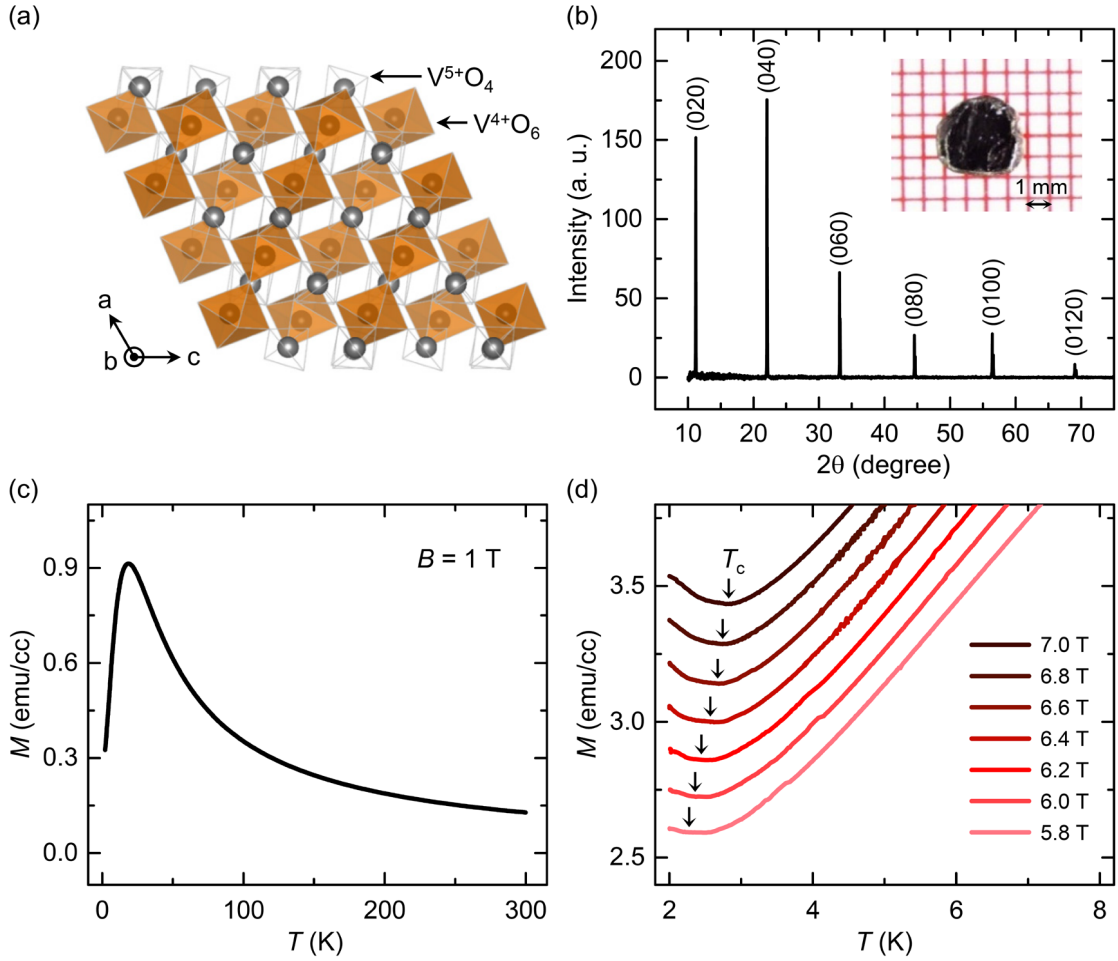

FIG. 1. Basic characterization of the single crystalline PVO. (a) Crystal structure of PVO (space group $C\bar{1}$). The $V^{4+}O_6$ octahedrons with $S = 1/2$ $V^{4+}$ ions construct the spin chains. (b) X-ray diffraction results of the single crystalline PVO. Inset: the optical image of a typical PVO bulk crystal. (c) Temperature dependence of the magnetization of PVO measured at $B = 1$ T. (d) Detailed temperature evolution of PVO's magnetization under various $B$ from 5.8 to 7.0 T. The arrows indicate the critical temperatures ($T_c$), below which an upturn of the magnetization is observed.

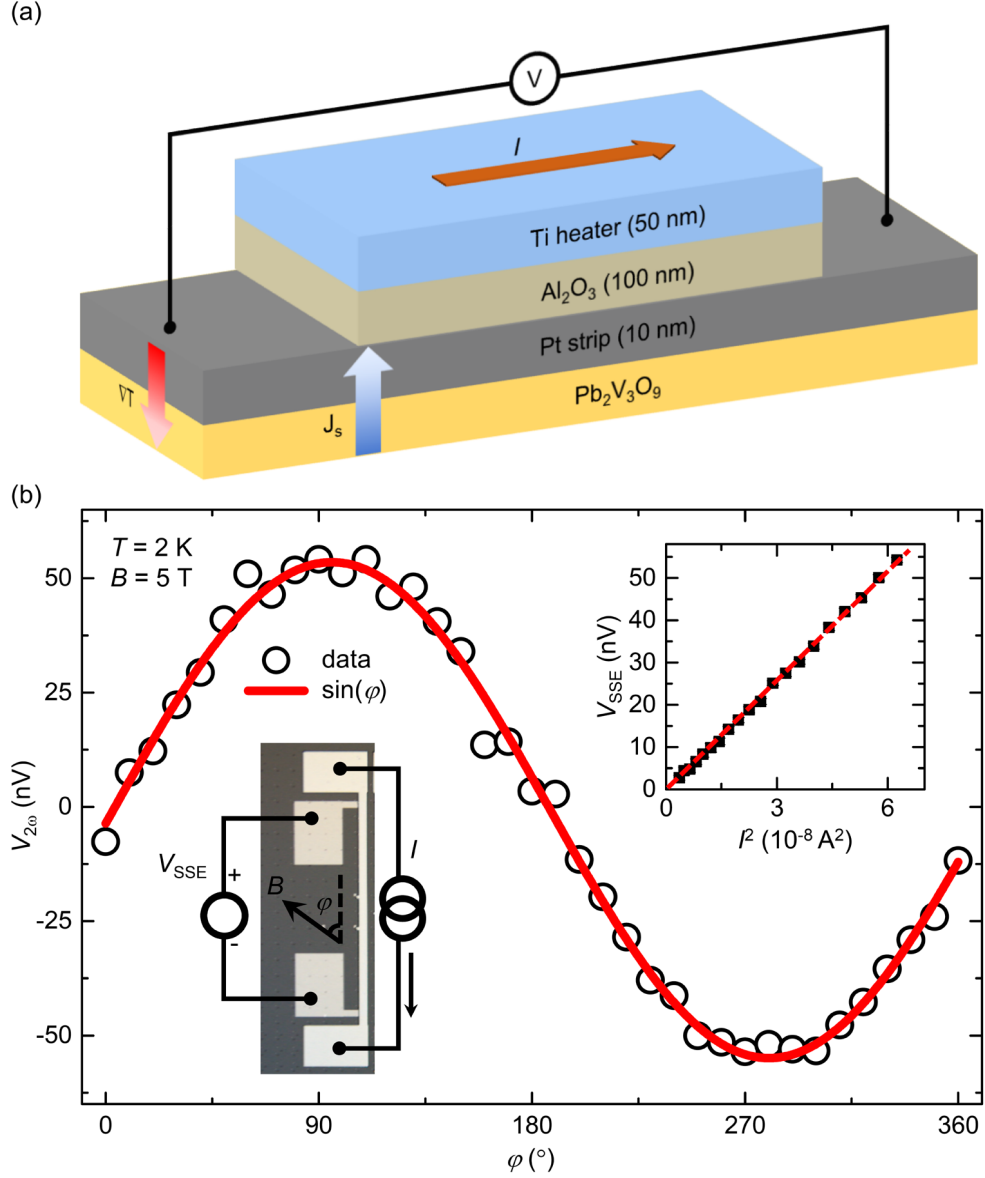

FIG. 2. SSE measurements of the single crystalline PVO. (a) Schematic of the SSE device structure and the measurement geometry. (b) $V_{2\omega}$ measured on the PVO device as a function of the in-plane magnetic field angle ($\varphi$) at $T$ = 2 K and $B$ = 5 T. Left inset: the optical image of the SSE device and the measurement geometry. $\varphi$ is the angle between the magnetic field and the Pt electrode. Right inset: $V_{SSE}$ as a function of the square of heating current ($I$).

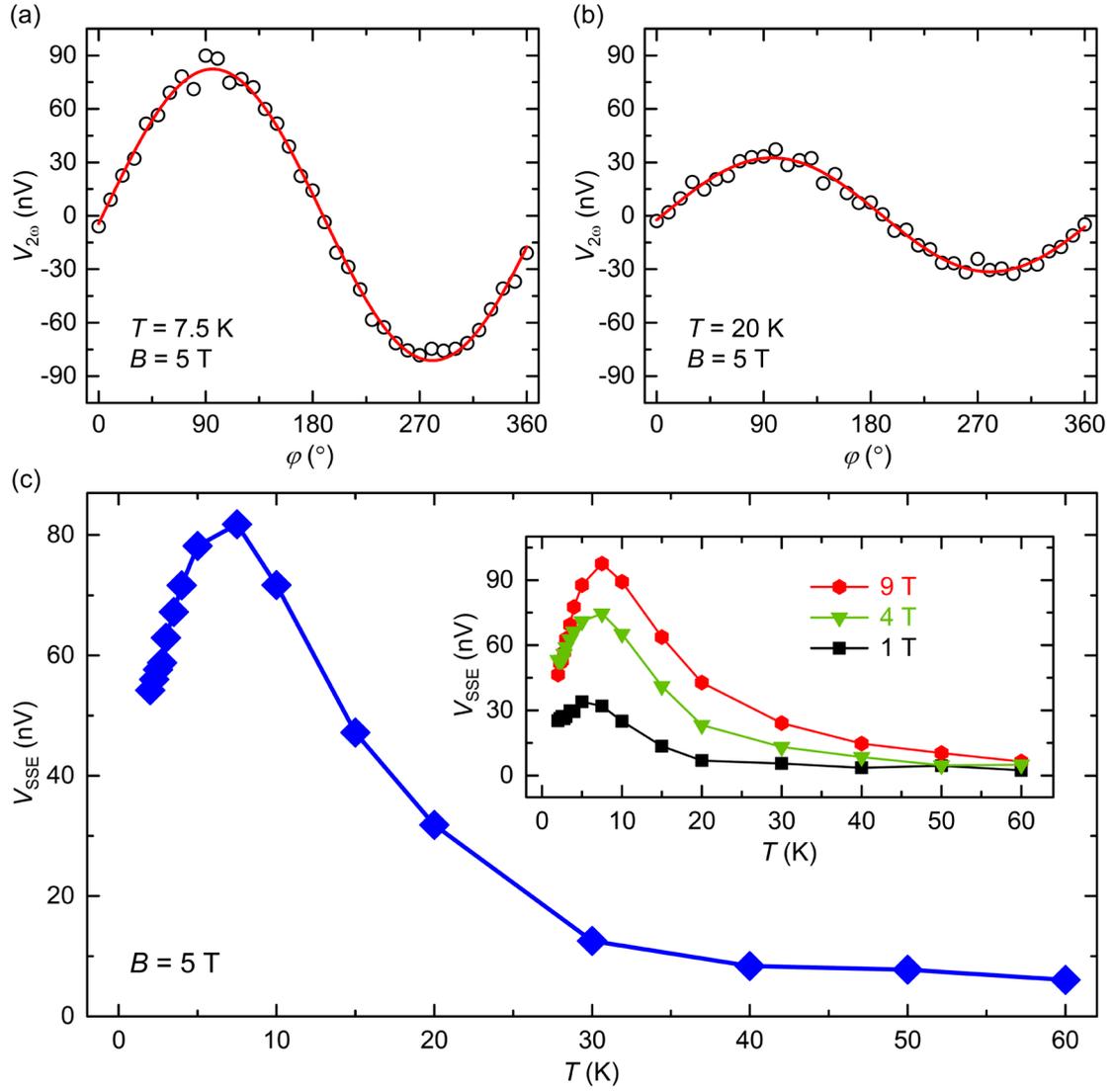

FIG. 3. Temperature dependence of the SSE of the single crystalline PVO. (a-b) $V_{2\omega}$ measured on the PVO device as a function of the in-plane magnetic field angle at $T = $ 7.5 and 20 K, respectively. (c) Temperature dependence of $V_{\text{SSE}}$ at $B = 5$ T. Inset: temperature dependence of $V_{\text{SSE}}$ at $B = 1$, 4, and 9 T, respectively.

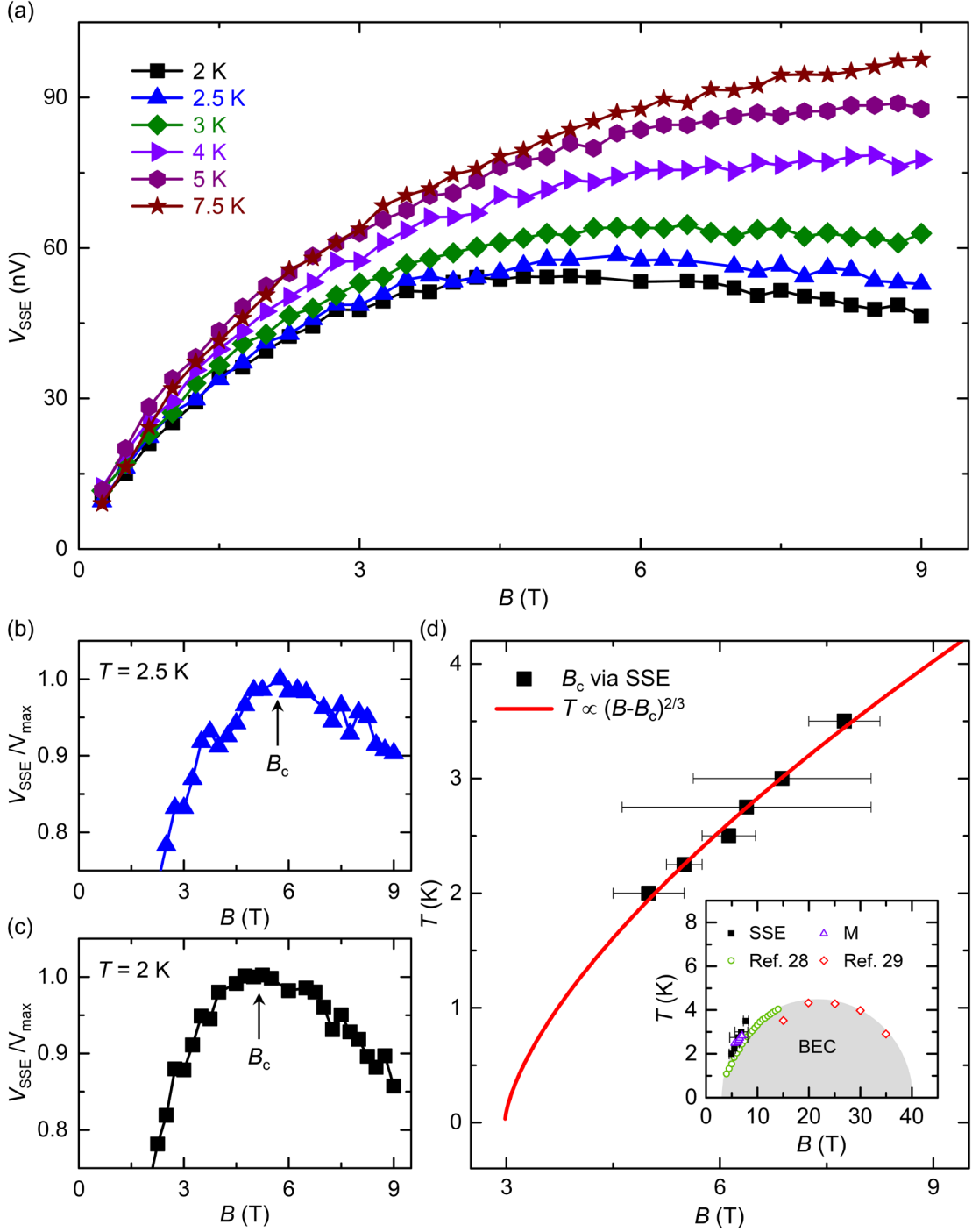

FIG. 4. Magnetic field dependence of the SSE of the single crystalline PVO. (a) Magnetic field dependence of $V_{SSE}$ at $T$ = 2, 2.5, 3, 4, 5 and 7.5 K, respectively. (b-c) Magnetic field dependence of the normalized $V_{SSE}$ at $T$ = 2.5 and 2 K, respectively. The arrows indicate the critical magnetic fields for the maximum $V_{SSE}$. (d) Phase boundary deduced from the temperature-dependent critical magnetic fields via SSE. Inset: The comparison of the phase diagram from SSE with that obtained from the present magnetization and pervious specific heat (Ref. 28) and torque (Ref. 29) measurements.

Inset in (d) reproduced with permission from J. Phys. Soc. Jpn. 80, 034710 (2011). Copyright 2011 the Physical Society of Japan. Inset in (d) reproduced with permission from Phys. Rev. B 81, 132401 (2010). Copyright 2010 the American Physical Society.